\documentclass[]{aa}
\usepackage{natbib}
\usepackage{graphicx}
\usepackage{txfonts}
\def\EBV{\mbox{E$_{\rm B-V}$}}

\def\AV{\mbox{A$_{\rm V}$}}
\def\AMSX{\mbox{A$_{\rm MSX}$}}
\def\RV{\mbox{R$_{\rm V}$}}

\def\HH{\mbox{H$_2$}}

\def\nH2{{\rm n}({\rm H}_2)}
\def\NH2{{\rm N}({\rm H}_2)}
\def\pccc{~{\rm cm}^{-3}} 
\def\pcc {~{\rm cm}^{-2}}

\def\Tstar#1 {\mbox{${\rm T}_{\rm #1}^*$}}
\def\Tsub#1 {\mbox{$T_{\rm #1}$}}
\def\TK  {\Tsub K }

\def\arcsec{\mbox{$^{\prime\prime}$}} \def\arcmin{\mbox{$^{\prime}$}}

\def\degr{\mbox{$^{\rm o}$}}

\def\p{\mbox{$^+$}}
\def\cotw {\mbox{$^{12}$CO}}
\def\coth {\mbox{$^{13}$CO}}

\def\h13cop{\mbox{{H$^{13}$CO\p}}}

\def\c3h2{\mbox{C$_3$H$_2$}}
 
 \def\R0{R$_0$}

\def\ddeg{{}^\circ\kern-.1em}  
\def\Msun{M$_{\rm sun}$}

\def\kms{\mbox{km\,s$^{-1}$}}

\def\E#1 {$10^{#1}$}
\def\E#1 {E{#1}}
\def\P#1,{$\nH2\TK~=~#1\times~10^4\pccc$~K}
\def\ec#1,#2,#3,{#1\,(#2)\E{#3}}

\def\H3{\mbox{H$_3$}}

\def\RH2{\mbox {R$_{\rm G}$}}
\def\fH2{\mbox {f$_{\HH}$}}
\def\FH2{\mbox {F$_{\HH}$}}
\def\W13{\mbox{W$_{13}$}}
\def\WCO{\mbox{W$_{\rm CO}$}}

\title{H II regions, infrared dark molecular clouds and the 
 local geometry of the Milky Way's nuclear star-forming ring 
 \thanks{Based on observations obtained with the ARO Kitt Peak 12m 
   telescope.}}
\author{H. S. Liszt\inst{1} }
\institute{National Radio Astronomy Observatory,
           520 Edgemont Road,
           Charlottesville, VA,
           USA 22903-2475}

\begin{document}
\date{received \today}
\offprints{H. S. Liszt}
\mail{hliszt@nrao.edu}
%

\abstract
   {}
    {To interpret the galactic center H II region complexes
   as constituents of a barred galaxy's nuclear star-forming ring.}
   {We compare 18cm VLA radiocontinuumm, $8-22\mu$ MSX IR and 
  2.6mm BTL and ARO12m CO emission in the inner few hundred pc.}
  {Galactic center H II regions are comparable in their IR appearance, 
  luminosity and SED to M17 or N!0, but the IR light distribution 
  is strongly modified by extinction at 8-22$\mu$, locally and 
  overall.  In Sgr B2 at $l > 0.6$\degr\ strong radio H II regions are 
  invisible in the IR.  In two favorable cases, extinction from individual 
  galactic center molecular clouds is shown to have $\tau \ga 1$ at 8-22$\mu$
  independent of wavelength.  The gas kinematics are mostly 
  rotational but with systematic $\pm 30-50$ \kms\ non-circular motion.
  Sgr B and C both show the same shell and high-velocity cap structure.}
 {The HII regions lie in a slightly-inclined ring of radius 
  $\approx$ 180 pc (1.2\degr) whose near side appears at higher latitude 
  and lower velocity and contains Sgr B.  Sgr C is on the far side and
   both Sgr B and C represent collisions with 
  material inflowing along the galactic dust lanes. Sgr E is a coincidental 
  aggregation of field objects seen tangent to the ring's outer edge. 
  Most of the volume interior to the ring is probably devoid of dense 
 gas and some emission seen at v=20-70 \kms\ toward Sgr A lies 
 outside it, in the ring}

\keywords{ interstellar medium -- molecules }
  

\titlerunning{The shiny ring hidden in the Milky Way Bar}

\maketitle

\section {Introduction.}

This is the third paper of a short series \citep{Lis06Shred,Lis08Multi} 
discussing the Milky Way's inner galaxy gas distribution in the context 
of the characteristic features which are present in nominally similar 
barred external galaxies \citep{ReyDow99,GreRey+99,RegShe+99} and in 
models of barred galaxy gas flow \citep{Fux99,RegShe+99,RegTeu04,RodCom08}.
Inasmuch as possible, but without being overly reductive, we wish to 
meld the local data, with its unique galactic perspective and overwhelming 
wealth of detail, with a framework which has proved so successful in
elucidating structure in other galaxies.   The desirability of 
doing this seems clear, and an entirely similar view has recently been 
adopted by \cite{RodCom+06} and \cite{RodCom08}.  However, connections
between Milky Way observations and those of other systems have been
obscured by the complexity and richness of the local observations and 
by our unique perspective within the galactic disk.  Some aspects of 
of the Milky Way have been made to appear unique simply by our unparalleled 
vantage point and common features in barred galaxies have been missed
in the Milky Way.

In the first two papers we discussed material associated
with the large-scale bar dust lanes.  We suggested that 
a series of enigmatic, highly-localized broad-lined molecular gas features 
should be associated with the shredding of molecular clouds at positions of 
gas uptake into the large-scale galactic dust lane shocks :  in other 
galaxies such phenomena are visible as broad lines and  velocity
shifts across the dust lane \citep{ReyDow99} but the individual uptake 
events and their surprising and quite exteme vertical elongation
 (common to all the Milky Way events)
are not discernible.  The peculiar 
kinematic signature commonly identified with the Milky Way's dust lane 
can be found in at least three distinct features in the northern hemisphere
(see also \cite{RodCom+06}) and the temporal sequence associated
with gas uptake and inward flow may perhaps be used to understand the vertical
course of the bar gas flow.

In this work, we focus on the structure interior to the inner ends 
of the prominent bar shock dust lanes, in the more immediate vicinity of 
the nucleus.  In other barred galaxies the star formation in this zone
occurs within narrowly-defined nuclear star-forming rings (NSFR; see 
\cite{KorCor04}). In the Milky
Way, this phenomenon is represented by the complex of sources 
Sgr A-E, the giant H II region-molecular cloud complexes at $|l| < 1.4$\degr\ 
corresponding to projected distance of some 200 pc from SgrA$^*$.  However, owing
to our viewing geometry, a ring morphology has been far from obvious; instead, 
the region has been generally characterized as the ``central molecular zone'' 
(CMZ) \citep{MorSer96} which in its most extreme form is described as being 
nearly filled with high density molecular gas.  Here we suggest that the CMZ 
is actually rather hollow, as is typically the case in barred galaxies having
prominent nuclear star-forming rings.

This paper considers the ring structure in detail, both locally within
individual H II region/molecular cloud complexes, and overall.  We compare 
the distributions and/or kinematics of three tracers, namely the L-band (18cm) 
radiocontinuum emission, 8-22$\mu$ IR emission and 3mm CO emission.
The organization of this work is as follows.  Observational material, some
of which is new, is discussed in Sect. 2.  Section 3 gives an overview of the
distribution of mid-IR and molecular emission in the inner $\pm$200 pc and
a detailed discussion of the individual off-nuclear H II region/molecular cloud 
complexes Sgr D, B, C, and E: in the most extreme case, the strongest
radio H II regions in Sgr B2 are lost to extinction in the IR, even at 22$\mu$.  
In Section 4 we derive the extinction originating locally in a few 
especially-prominent infrared-dark (molecular) clouds (IRDC): consistent 
with previous results \citep{IndMat+05,NisNag+08} 
and a plateau in the model interstellar extinction curves of \cite{WeiDra01}, 
the extinction shows no decline over the MSX bands.  In Sect. 5 we derive 
IR-radio spectra of the Sgr source complexes and their compact constituents, 
showing how the IR-radio continuum comparison can be used to infer high 
extinction when it is not obvious in the underlying source morphology.
In Sect. 6 we discuss the global placement of sources in a putative
ring, within which the non-circular motion and front-back inclination are just
large enough to distinguish the near and far ring segments.  Section 7
is a summary.  Note that the Sun-center distance is taken as 
R$_0$ = 8 kpc so that 1\arcmin] corresponds to 2.33 pc.

\section{Observational material considered}

\subsection{New observations}

The observational material considered here comes largely from published
sources, but it also includes a substantial component of previously 
unpublished large-scale mapping of J=1-0 \coth\ emission from the 
present ARO (former NRAO) 12m telescope on Kitt Peak, observed in 1995 May
 by the author.  
Maps were made in the vicinity of Sgr B and D at $0.5\degr 
< l < 1.2 \degr$ and around Sgr E at $l=-1.2$\degr.  These are
fully spatially-sampled data with 1\arcmin\ spatial resolution, 1 MHz
(2.7 \kms) velocity resolution and typical rms noise 0.1 K.

\begin{figure*}
\includegraphics[height=16.9cm]{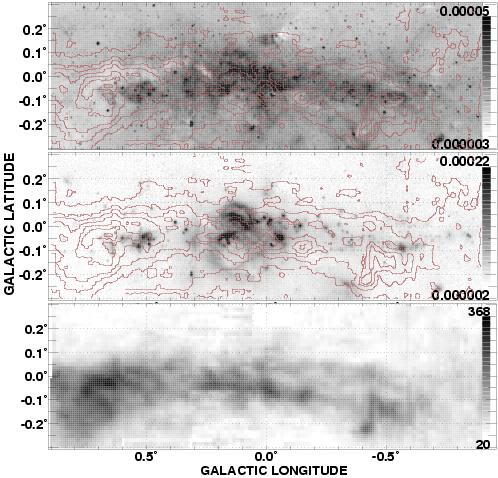}
\caption[]{Mid-infrared continuum and molecular gas in the innermost
$\pm$125 pc (R$_0=8$ kpc) of the Milky Way.  Top:  MSX emission in Band A 
($8.3\mu$).  Middle: MSX emission in Band E ($21.8\mu$).  Bottom and superposed 
as contours in the middle panel: \coth\ emission integegrated at
$-100 <$ v $< 100$ \kms\ from the survey of \cite{BalSta+87}.}
\end{figure*}

\begin{figure*}
\includegraphics[height=8.8cm]{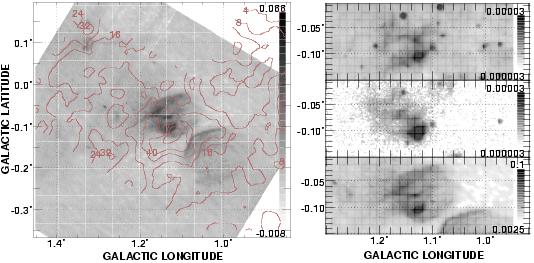}
\caption[]{Views of the Sgr D region.  Left:  VLA radiocontinuum emission
at 1616.4 MHz (grayscale) and overlayed contours of \coth\ emission 
integrated over $-20 <$ v $< 0$ \kms\ as mapped at the then-NRAO, now-ARO 
12m telescope with 1\arcmin\ resolution.  Right:  Sgr D as seen at MSX Band
A (top), Band E (middle) and at 1616.4 MHz.}
\end{figure*}

\subsection{Previously published CO emission line data}

Also shown here are maps drawn from the BTL galactic center
CO survey \citep{BalSta+87} at 1.7\arcmin\ resolution.  
For \cotw\ these data are on a 1\arcmin\ grid with 
1 MHz or 2.6 \kms\ velocity resolution and typical rms 0.5 K.
For \coth\ the grid spacing is 30\arcsec\ and the typical rms
noise level is 0.15 K.

\cite{LisSpi95} published 45\arcsec\ resolution maps of molecular gas near Sgr C
from the SEST telescope, and these data are used here in Fig. 5.

\subsection{Nominal CO-\HH\ conversion}

Unless otherwise noted, velocity-integrated intensities of \cotw\ and 
\coth\ in units of K-\kms, denoted respectively by \WCO\ and \W13, 
are converted to 
\HH\ column density N(\HH) assuming a typical galactic disk 
conversion N(\HH) $= 2\times10^{20}$ \HH\ $\pcc$ \WCO, and 
\WCO/\W13\ = 6 which is typical of galactic center material
in the H II region complexes in our data and the BTL survey.

\subsection{VLA L-band continuum}

This is the data discussed originally by \cite{Lis92} 
and \cite{LisSpi95}.  It was taken in a 12.5 MHz band centered at 1616.4 MHz 
(a region of the spectrum subsequently lost to use by the Iridium satellite 
constellation) in the VLA C and D configurations and has spatial resolution 
 13\arcsec $\times$ 23\arcsec.

\subsection{Radio recombination line data}

The data, shown here in Fig Y, were taken in 1990 at the NRAO 43m
telescope in Green Bank in 1990, and were first shown by \cite{Lis92}.

\subsection{MSX maps}

The Mid-Course Space Experiment (MSX; \cite{PriEga+01}) mapped 
the galactic center region 
at 21\arcsec\ resolution on a 6\arcsec/pixel grid in four bands at
center wavelengths ranging from $8\mu$ to $22\mu$.  These data
are provided as specific intensity in units of W (m$^{2}$-sr)$^{-1}$. 
These can be expressed in Jy at individual pixels using the 
conversion factors given in Table 
1 corresponding to the solid angle of 6\arcsec\ pixels and the
the stated photometric bandwidths of the MSX spectral bands expressed in
Hz.  Figure A.1 in Appendix A illustrates the 50\%-response MSX bandwidths
graphically on curves of the extinction cross-section per H atom for
galactic dust \citep{WeiDra01}. 

\begin{table}
\caption[]{MSX Bands and conversion from pixel values to Jy}
{
\begin{tabular}{lcc}
\hline
Band & wavelength & X$^1$ \\
      &$\mu$  & sr Hz$^{-1}$\\
\hline
A & 8.28 & 5.53$\times10^{3}$ \\
C & 12.13 & 2.40$\times10^{4}$ \\
D & 14.65 & 2.66$\times10^{4}$ \\
E & 21.85 & 2.02$\times10^{4}$ \\
\hline
\end{tabular}}
\\
$^1$ X is the factor by which MSX pixel values or
sums of pixel values should be multiplied to 
convert to Jy. 
\\
\end{table}

\subsection{A word about the figures used in this version of the paper}

The postscript versions of several of the figures created for this paper are
intractably slow to load on-screen and all the figures have been replaced here 
by serviceable, more compact versions suitable for pdflatex, which will not 
be used in the published paper.  The striations in Fig. 2 and 11 are 
artifacts of image conversion and will not appear when published.
 
\section{Structure of the inner region and individual sources}

The top two panels of Fig 1 show the mid-IR brightness in the MSX A and 
E-bands (around 8.3 and 21.8 $\mu$, respectively, see Table 1 and 
Fig. A.1).  Superposed in each case are contours of the velocity-integrated 
intensity of \coth, \W13, at $|{\rm v}| \le 100 $ \kms\ from the survey of 
\cite{BalSta+87}.  Shown at bottom in grayscale is the same distribution 
of \W13.  The 0.9\degr\ radius which contains most of the IR brightness 
projects to 125.6 pc across the line of sight at a Sun-center distance of 
8 kpc (1\arcmin\ corresponding to 2.33pc).  The outlying sources Sgr D 
and Sgr E at $|l| > 1.15$\degr\ 
($> 160$ pc) which define the outer edge of the nuclear star-forming 
region are not shown in Fig. 1 but will be discussed separately in 
this Section along with the other sources.

It is clear from observations of galactic H II region complexes 
(see \cite{PovSto+07,WatPov+08} and Appendex C) that the shorter 
wavelength panel in the middle of Fig. 1 mainly shows very hot dust 
interior to the ionized region so that the Sgr B, A, and C H II regions 
at 0.7\degr, 0.2\degr\ and -0.5 \degr, respectively, appear with 
greater clarity.  The longer wavelength panel at the top of Fig. 1
shows more broadly-distributed material within the ambient neutral gas, 
and especially (as discussed below) at the H II region/molecular gas 
interfaces.

All of the Sgr H II regions lie 
in a rather thin band running parallel to the galactic equator between 
b $= -0.05$\degr\ (the latitude of Sgr A$^*$) and b $= -0.10$\degr).
However, the large-scale distributions of molecular gas and IR light are 
noticeably bowed, with most of the emission occuring well below the 
galactic plane at the extremes in Fig. 1.  The IR distribution in the 
uppermost panel at 8$\mu$ is clearly affected by infrared dark clouds 
visible in CO, especially in an extended distribution of 
foreground extinction at l $> 0.25\degr$ shrouding Sgr B.  The downward 
slant of the upper edge of the 8$\mu$ emission at postive longitude 
in Fig. 1 at top is more obviously an artifact of this extinction,
but extinction is also visible, albeit somewhat more faintly, at 
the northern edge of  the 8$\mu$ light distribution at negative 
longtitude where it also bows downward. It is straightforward to
visualize how the accumulation of absorption near the ends of
a ringlike distribution might be responsible for this, when 
the ring is slightly inclined so that the nearer portions are
seen at higher latitude.  However, much of the extinction is 
actually indigenous to the same H II region complexes which are
responsible for the IR light in the first place. 
 
Extinction in the Sgr B complex around l = 0.6\degr\ largely but 
not exclusively (see Sect. 3.2) arises in molecular gas at 0 
\kms\ $<$ v $<$ 50 \kms\ and one especially clear case of extinction 
by this gas is seen just above the equator between Sgr A and B at  
l $\approx 0.25\degr$.  Another prominent extinction arising in
gas at negative velocity appears just above the Sgr C H II region at
 l $= 359.5\degr$ and both of these cases are discussed in detail
in Sect. 4.  It is less obvious that emission at the longest
wavelength IR map in the middle panel is significantly
affected but the extinction of the individual IRDC shows no 
obvious wavelength dependence from 8-22$\mu$ and some foreground
extinction at 22$\mu$ which is not apparent from the morphology
may be inferred from the spectra which are discussed in Sect. 5.

The general question of how the material represented in Fig. 1 
is actually distributed in space is left for Sect. 6 
but the appearance of the molecular gas at bottom is clearly rather
looped and somewhat hollow at negative longtiudes.  At positive longitudes
the molecular gas in the Sgr B complex is so bright and so extended 
that the z-height of the material overall is otherwise obscured.  
The expectation is that the star forming regions in a strongly-barred 
galaxy like the Milky Way will be situated in a radially-narrow,
approximately circular,  mostly-rotating ring and the inference of 
a ring geometry for the Milky Way gas is discussed in Sect. 6.  
The nuclear region in our Galaxy is the only one whose
vertical structure is distinguishable and, at least at negative
longitudes, it appears that a slight scalloping may have displaced
the front and back gas portions, so that they are not projected 
on top of each other.  In Sect. 6 we also discuss how the front and back
ring portions are distinguishable kinematically, owing to departures 
from pure circular motion.

The structure of the individual H II region/molecular cloud complexes is
discussed in the following subsections, progressing from highest (Sgr D) 
to lowest (Sgr E) galactic longitude.


\begin{figure}
\includegraphics[height=10cm]{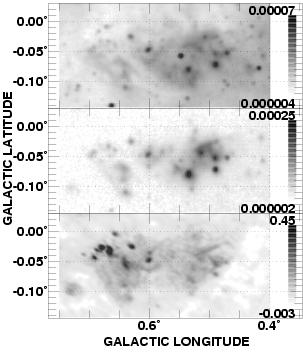}
\caption[] {Sgr B as seen at MSX Band A (top), Band E 
(middle) and at 1616.4 MHz.}
\end{figure}

\subsection{Sgr D}

The radiocontinuum properties of  Sgr D were described by \cite{Lis92} and
\cite{MehGos+98} and a Band E MSX image appears in \cite{ConCro04}.  
The source is visible at $\lambda2\mu$ \citep{BluDam99}.  Although
it lies at the ``right'' galactic latitude, its location within the
ring is contentious.

Figure 2 shows the source morphology in the radio and MSX IR bands.  
Sgr D consists of an H II region to the North, visible at both radio and 
infrared wavebands, and an adjacent shell radio supernova remnant (radio 
SNR) which appears only in the radio (see also \cite{Gra94}).  The H II region 
 is associated with a peak in the \coth\ emission having \W13\ 
$\approx 50$ K \kms\ and line brightness 5 K at v $\approx -17$ \kms\ (see 
also Fig. 2) near the recombination line velocity (Fig B.1). The nominal 
mass of associated molecular material (see Sect. 2.2) is 
$\approx 2\times10^5$ \Msun.  

The bright \coth\ peak most likely represents a region of interaction between 
the ionized and neutral gas distributions; overall the molecular gas has a 
shell-like appearance which cradles the southern portions of the extended 
thermal radiocontinuum distribution.  Portions of this shell and the wishbone 
structure to the north are visible in the MSX Band A 
emission at top right in Fig. 2.  Longer wavelength IR emission in MSX Band E 
is more nearly dominated by hot material in the compact H II region but extended 
structure is also visible, especially to the north.  Although it is not
obvious that the IR source morphology is affected by extinction, the global 
spectrum shows some deficiency of IR emission compared to the 
radiocontinuum, which in other cases results from  foreground extinction 
of the IR (see Sect. 5).

The  placement of Sgr D along the line of sight remains somewhat of a puzzle 
owing to its recombination line velocity of -9 \kms\ (see Fig.B.1).  
Interpretation of absorption measurements has placed it at \citep{ConCro04} 
or beyond \citep{Lis91,MehGos+98} the galactic center. In particular, 
\cite{Lis91} argued for a position outside the bulge based on the narrowness 
of the mm-wave emission lines from the nearby molecular gas.  Examination
of molecular emission over a larger region does not provide much context 
for placement of Sgr D, whose velocity is distinct from those of other 
similarly-narrow but spatially-extended ridges of emission in the
region.  Owing to the extraordinary kinematic complexity of the region, 
which falls between the nuclear ring and the innermost uptake event
into the dust lane at l=1.3\degr\ \citep{Lis06Shred}, there actually is gas 
at rather low $|v|$ which is undeniably part of the inner-galaxy gas.

Within the ring it is expected that stars and natal gas will separate 
and phase mix with other ring members after the stars form, making it less 
likely 
that the evolved progenitor of the Sgr D SNR and the O-stars in the H II 
region actually formed from the same cloud.  Moreover, it is possible that 
the SNR is an interloper in the vicinity
of the H II region and molecular cloud (if all are actually sited within
the ring), so that proximity still does not demand a physical association.  
However, if an association between the H II region and SNR could be 
demonstrated, it would have interesting consequences for star formation 
mechanisms within the ring gas.  A physical association between the H II 
region and SNR could be investigated by searching for 1720 MHz OH maser 
emission around the SNR.

\begin{figure*}
\includegraphics[height=6.7cm]{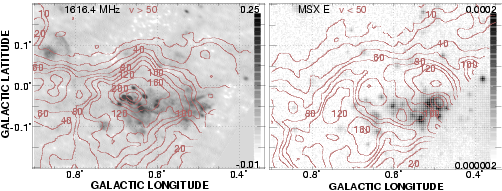}
\caption[]{Sgr B at infrared and radio wavelengths. Left: grayscale
of 1616.4 MHz radiocontinuum, overlayed by contours of \coth\
at 1\arcmin\ resolution integrated at $50 <$ v $< 100$ \kms\ which 
includes the radio
recombination line velocites (see Fig. 3).  Right:  MSX Band E emission
in grayscale with superposed contours of \coth\ integrated emission
at $10 <$ v $< 50$ \kms, which forms a shell within which most of the
Sgr B radio and IR continuum is contained.  Note the strong absorption 
associated with the molecular gas at right and the near total absence 
of Sgr B2 at $l > 0.6\degr$ even at 22$\mu$ in Fig. 3.}
\end{figure*}

%
%

\subsection{Sgr B}

The presence of extinction in the Sgr B source complex around l=0.6\degr\
is apparent in Fig. 1 but its full influence can better be seen by comparing 
source structure in the radio and IR continuum as in Fig. 3.  Simply put, 
the strongest H II regions visible in the radio (at $l \ga 0.6$\degr) 
are mostly absent at both 8$\mu$ and 22$\mu$. Although the physical
structure of Sgr B arises from an interaction between the neutral and 
ionized gas (which partly overlap in velocity and are physically co-mingled), 
the absence of the strong radio H II regions in the IR can only result 
from near-complete attenuation of the IR flux.

The radiocontinuum and MSX Band E images are shown in Fig. 4 along with
the integrated \coth\ emission at velocities above and below 50 \kms.
The higher velocity emission, which overlaps the recombination line
velocities (62 \kms; see Fig. B.1), is shown over the radiocontinuum 
with its clear
image of the H II regions: lower velocity emission, from more extended
gas responsible for the IR extinction, is overlaid on the MSX Band E image 
in which the extinction is the most important aspect of the extended neutral
gas distribution.  The IR and radio continuum emission largely exists inside
a cavity in the lower-velocity molecular gas, which is partly filled in by
material at the higher velocity.  The brightest compact radio H II regions
which are so heavily extincted are seen projected squarely against very bright 
CO emission at comparable velocity, and must be behind the molecular gas.

The low-velocity shell and high-velocity cap have been interpreted as a 
cloud-collision in a series of papers by Hasegawa and his collaborators 
\citep{HasAra+08,SatHas+00,HasSat+94}; this explanation accounts for 
many of the properties of the regions chemistry and masers.  Sgr B2 can 
probably be identified with the contact region between the star-forming
ring and gas inflowing in the dust lanes in the nuclear regions of the Milky
Way Bar (e.g. \cite{RegTeu03} and Fig. 9 of \cite{RegShe+99} or recent
discussions in \cite{Lis06Shred} and \cite{RodCom08}).  A similar spatial
and kinematic shell and cap distribution of the CO emission is also seen 
in Sgr C (see immediately below).

\subsection{Sgr C}

Although far less active than Sgr B, the Sgr C H II region near l = 359.45\degr, 
b = -0.1\degr, at  v = -60 \kms\ (see Fig. B.1) is positioned nearly 
symmetrically in space and velocity.  The radiocontinuum structure has 
been discussed by \cite{Lis85}, \cite{TsuKob+91} and \cite{LisSpi95} and 
maps of the molecular gas were discussed by \cite{LisSpi95} and 
\cite{StaStu+98}.


The IR and radiocontinuum structure of Sgr C is shown in Fig 5.  In the 
radio,  Sgr C consists of extended, round, thermal emission from an H II 
region powered by one or a small number of O-stars \cite{LisSpi95} and 
an adjacent,elongated, non-thermal filament which was the first such 
object to be seen outside Sgr A.  Partly based on the structure in Sgr C 
\cite{Yus03} argued that non-thermal filaments in the galactic center 
are caused by interaction of young stellar clusters and ambient dense 
gas, but \cite{Roy03} argued that the H II region and filament are 
separated along the line of site and not physically related, owing 
to differences in their H I absorption spectra.

Based on the limited \coth\ mapping shown in Fig. 5 \cite{LisSpi95} noted
merely that the Sgr C H II region is placed between the two molecular emission
distributions at -125 $<$ v $<$ -85 \kms\ and -85  $<$ v $<$ -45 \kms.
It has been something of a puzzle why gas at such widely separated
velocities should all brighten near Sgr C, but larger-scale imaging of 
the molecular gas (Fig. 6) shows that the kinematic structure is 
actually very like that in Sgr B: a smaller core of emission near 
the recombination line velocity (-65 \kms; Fig. B.1) is surrounded 
by a partial ring of gas which is displaced in velocity.  Apparently,
Sgr C is actually physically associated with both kinematic components
of the gas and, like Sgr B, it may represent the interaction of inflowing
material in the dust lane with pre-existing material in the SFR.

MSX Band E emission coincides with the radio H II region and
arises from heated material within it.  The more extended 
and extincted MSX Band A emission occurs
between the blue-shifted gas to the North and
the cloud at the H II region velocity; only a rather minor 
portion of the MSX Band A light overlaps the H II region.  
As such it might be  associated with 
either of the two kinematic components. Although there is weaker emission
at the H II region velocity across the prominent Band A extinction
north of Sgr C, the CO emission around -100 \kms\ is much stronger and
similar in appearance to the region of IR extinction.  
The extinction likely arises in the lower
velocity gas (see just below and Fig. 6). As noted by \cite{Roy03} 
there is H I absorption at -100 \kms\ toward the H II region.
The extinction is discussed in more detail in Sect. 4.

\subsection{Sgr E}

The radiocontinuum structure and recombination line kinematics of Sgr E 
around $l = -1.3$\degr, $b = -0.1$\degr, v = -215 \kms\ (Fig. B.1) have 
been observed by 
\cite{Lis92}, \cite{GraWhi93} and \cite{CraCla+96}.  At a maximum
projected radius of 195 pc corresponding to  l = 358.6\degr, Sgr E 
marks the outer boundary of the galactic center star-forming ring in both 
galactocentric radius and velocity.  Sgr E lacks a symmetric counterpart 
at positive longitude, especially as regards the velocity.

Sgr E is unusual among the named galactic center H II region complexes in 
being devoid of extended emission.  The very high observed velocities imply 
that 
the sources are observed near the sub-central point, in which case
the line of sight velocity gradient is small and the sources could be 
distributed over a long path where the line of sight is tangent to the 
edge of the ring.  Observing Sgr E may afford the opportunity to study a 
sample of somewhat older (i.e. B-type) ``field'' stars which have phase-mixed 
around the ring but the Sgr E source complex is not a physical entity.

The IR and radiocontinuum structure of Sgr E is shown in Fig. 7.  Nearly 
all of the radiocontinuum sources have IR counterparts.  The radiocontinuum 
source at l = 358.6\degr, b = 0.06\degr\ lacks such a counterpart and
was singled out by \cite{CraCla+96} for lacking a radio recombination line:
It is probably an extragalactic interloper and perhaps a useful source
for acquiring comparison absorption spectra through the galactic disk.

The nominal total \HH\ mass represented in Fig. 7 is $1.1\times10^6$ \Msun\ 
(see Sect. 2.2).

\section{IRDC and the CO-\HH\ conversion factor}

Extinction of the IR radiation by darker foreground material is apparent 
in Fig. 1 and two especially well-defined cases of association between 
molecular emission and extinction in the MSX bands are illustrated in Fig.
8 and 9.  

\subsection{The IRDC near Sgr C}

A very prominent IRDC seen just north of Sgr C was noted in 
Sect. 3.3 and illustrated in Fig. 5.  It probably originates in
gas seen in CO emission at -135 $<$ v $<$ -85 \kms\ (also see
Fig. 6).   As seen in Fig. 6 the peak CO brightness associated 
with the IRDC near Sgr C is \WCO\ = 480 K \kms.  Fig. 8 at top 
shows a latitude cut across the IRDC 
in MSX Band A.  The peak absorption optical depth is 
straightforwardly estimated as 0.75, corresponding to an 
extinction of 0.81 mag.  Extinction is probably present at 
22$\mu$, illustrated by the transverse cut in the lower panel 
of Fig. 8 which shows a clear dip at the expected 
latitude.  The extinction at 
both wavelengths must be comparable but determination of the 
spectrum of the extinction is better left for the 
even better-defined extinction between Sgr A and Sgr B. 

\subsection{The prominent IRDC between Sgr A and Sgr B}

MSX fluxes over a spatial cut across the especially prominent 
feature between Sgr A and Sgr B around l = 0.25\degr\ in Fig. 1 are shown
in Fig. 9.  The path of this cut runs at a 45\degr\ angle with respect
to the galactic equator, along the short axis of the absorption, from 
l = 0.4\degr, b=-0.1\degr\ to l = 0.2\degr, b=+0.1\degr.   Shown at top
is a strip across the full extent of the cut in MSX Band A, illustrating
how a baseline was established to gauge the strength of the absorption.
At bottom in Fig. 9 are the absorption and optical depth (inset) profiles which
ensue.  They are lower limits because we have assumed that the absorbing 
material is black and we have not accounted for contamination by unrelated 
foreground emission.

\begin{figure}
\includegraphics[height=9.24cm]{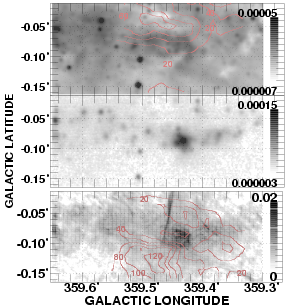}
\caption[]{Sgr C.  Top: MSX Band A emission with overlaid contours
of emission at $-135 <$ v $< -85 $ \kms\ at 1\arcmin\ resolution 
showing the higher-latitude,
lower-velocity branch of the  molecular ring (see Figs 1 and 2). 
Middle:  MSX Band E.  Bottom 1616.4 MHz radiocontinuum (grayscale) 
with overlaid contours of \coth\ emission at $-85 <$ v $< -45 $ \kms, 
which includes the recombination line velocity of the Sgr C H II region 
(see Fig. 3). }
\end{figure}

\begin{figure}
\includegraphics[height=9.6cm]{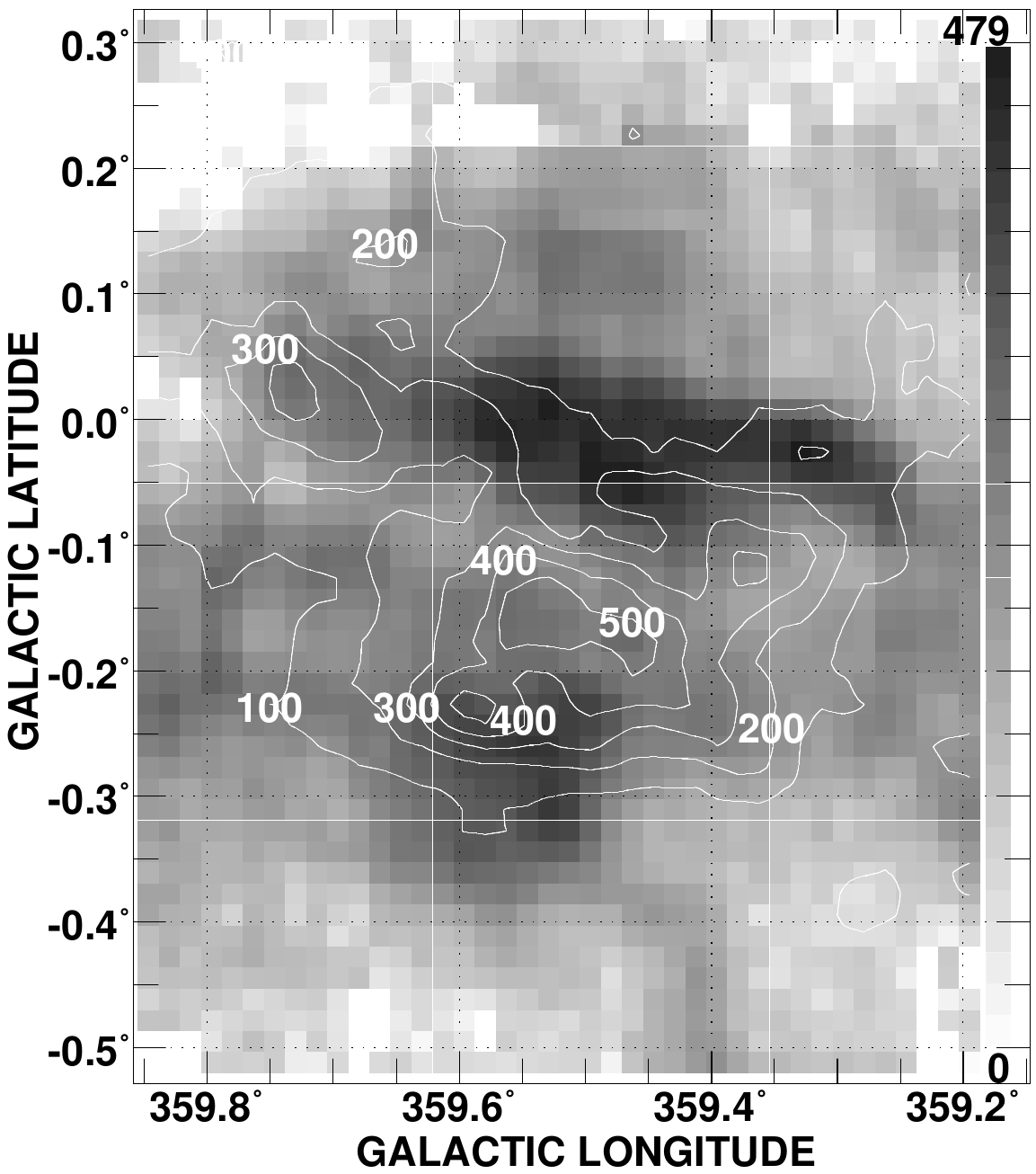}
\caption[]{\cotw\ emission near Sgr C.  The grayscale represents 
emission at 1.7\arcmin\ resolution integrated over the range 
$-135 <$ v $< -85 $ \kms\ 
and the contours are for emission integrated over the range 
$-85 <$ v $< -35 $ \kms\ which includes the recombination line velocity
of Sgr C as shown in Fig.B.1}
\end{figure}

The extinction is rather gray, with no obvious variation with band 
wavelength and a peak optical depth of 1.3 in all bands, corresponding 
to extinction of 1.4 mag.  Wavelength-independent extinction in 
this region of the spectrum has recently been noted by \cite{IndMat+05}
and \cite{NisNag+08} and is broadly consistent with a plateau in the
interstellar extinction in the dust models of \cite{WeiDra01}.
The overlap of the MSX 50\% photometric 
bandwidths with the wavelength-dependent extinction coefficients of 
\cite{WeiDra01} is shown in Fig. A1.  

At least approximately, an extinction coefficient 
C$^{\prime}$/H = $1.3\times10^{-23}\pcc/$H  can be attributed 
across the MSX A-E bands.  This implies column densities 
of $0.75/{\rm C}^{\prime} = 6\times10^{22}\pcc$ and
 $1.3/{\rm C}^{\prime} = 1.0\times10^{23}\pcc$ for the IRDC
near Sgr C and around l = 0.25\degr, respectively, given their 
optical depths of 0.75 and 1.3.  These then correspond 
to visual extinction \AV\ = $ 6\times10^{22}\RV/5.8\times10^{21} $
= 32.1 (\RV/3.1) mag and  53.4 (\RV/3.1) mag.  
Alternatively, \AV/\AMSX = 39.4 (\RV/3.1) given the
common assumption that \EBV/mag = N(H)$/5.8\times10^{21} \pcc$ 
\citep{SavDra+77}.

As suggested by the images Fig. 1, more detailed images of CO emission 
(not  shown here) indicate clearly that the material responsible for the 
absorption at l=0.25\degr\ is at 0-50 \kms, i.e. it is the same 
material responsible for the extinction around the Sgr B complex as 
discussed above in Sect. 3.2.  At the deepest trough in the IRDC, 
the peak integrated CO brightness in the 1.7\arcmin\ resolution BTL 
survey is 505 K \kms,  which is some 400 K \kms\ higher than at 
adjacent positions outside the IRDC.  This is equivalent to an 
CO-\HH\ conversion factor 
N(\HH)/\WCO\ = 0.5 N(H)/\WCO\ = $1.25\times10^{20}$ H (\RV/3.1)
$\pcc$. A consistent result is also obtained for the feature near
Sgr C.  At this level of accuracy the CO-\HH\ conversion factor
for galactic center gas is at most modestly below that for material
near the Sun. 

\begin{figure}
\includegraphics[height=12.5cm]{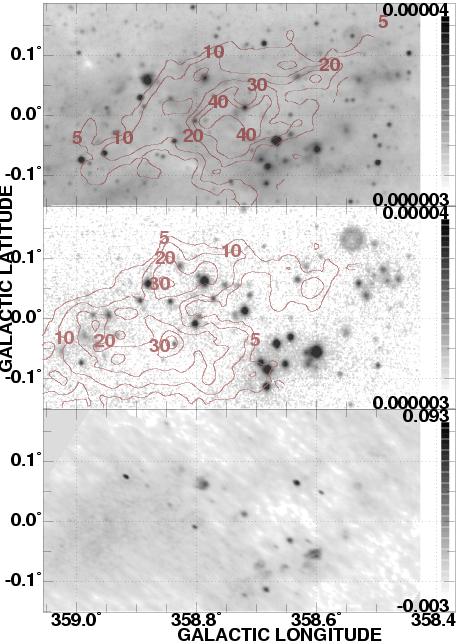}
\caption[]{Views of Sgr E.  Top: MSX Band A overlaid with contours 
of \coth\ emission at 1\arcmin\ resolution, 
integrated over $-225 <$ v $< -200$ \kms.  
Middle:  MSX Band E (grayscale) and \coth\ emission integrated 
over  $-200 <$ v $< -175$ \kms.  The recombination line velocities
of the radio-brighter sources at l = 358.6\degr\ and 358.8 \degr\
are -210 \kms and -190 \kms, respectively.} 
\end{figure}

\subsection{Sgr B}

The IR extinction in Sgr B is so very spatially extended that
a reliable baseline flux measurement did not seem feasible.  The integrated
CO intensities for the lower-velocity gas responsible for the IR extinction
are shown in Fig. 6 and Fig. 7 and are somewhat larger than are
present in the two IRDC just discussed, 140-160 K \kms\ for \coth\ and 
800 - 1000 K \kms\ for \cotw.  As discussed in Sect. 6, comparison of 
radio and IR fluxes suggests that the very strongest radio H II regions 
in the eastern part of Sgr B are behind material with optical depth of 
order 4 in the MSX bands.

\section{Global and compact H II region source spectra}

Figure 10 shows IR-radio spectra of many sources in the galactic center
region, extracted from the MSX and 18cm VLA maps.  All of the radiocontinuum
fluxes have been scaled upward by a uniform factor of 250 to allow convenient
viewing.  The fluxes shown are simple pixel sums over identical regions in
all bands (for each source), without correction for the background.

Also shown in Fig. 10 are comparison spectra for two galactic H II regions
outside the center, namely, M17 \citep{PovSto+07} and N10 \citep{WatPov+08}.
IR fluxes for these objects were also extracted from the MSX maps but the radio
fluxes are derived from the MAGPIS survey \citep{HelBec+06}, 
scaled by the frequency dependence of optically-thin free-free emission
to account for the slight difference in frequency (1390 vs. 1618 MHz).
Except at the upper left, the overall flux scale of the M17 spectrum
shown in the various panels of Fig. 10 is arbitrary, because only a
comparison of slopes is needed.  At the upper left, the M17 fluxes are
absolute but have been rescaled to 8 kpc distance (from 1.6 kpc) and 
(as noted there) further divided by a factor of two to facilitate comparison. 
The spectrum of N10, a galactic H II region bubble at a distance of 4.9 kpc, 
is also shown at upper left, without rescaling.  The morphology of N10 is 
generic; it strongly resembles Sgr C (see Fig. C1 and Appendix C).

In the following discussion, the behavior of the larger and more compact
sources is discussed separately.  Although the Sgr B, C, and D regions 
as a whole strongly resemble M17 and N10, allowing for the straightforward
possiblity of some additional extinction in the galactic center sources
in the IR bands, the spectra of compact sources are coherent within 
indivdiual complexes but rather different from one to another, especially 
between the two halves of Sgr B.

\subsection{Extended sources}

Integrated spectra of the extended galactic center regions Sgr B, C, and D are 
shown at upper left; they resemble those of M17 and N10 in form and bracket 
them in absolute flux.  This establishes that the galactic center sources are 
very much like galactic disk H II regions, as long as extinction in the IR (which
must be higher for the nuclear ring sources) is not too 
strongly wavelength dependent from 8-22$\mu$.   However, in the case where the foreground 
extinction is obviously the greatest, Sgr B at l $>$ 0.6\degr, the radiocontinuum 
flux is conspicuously large (by a factor of five) relative to M17, suggesting that 
the great majority of the total IR flux of the eastern half of Sgr B is absent
at/below $22\mu$.  Sgr D also has a somewhat large radio/IR flux ratio, but
somewhat less in comparison with N10, than with M17.  The slight difference in 
IR/radio flux ratios between M17 and N10 could represent normal variation in
the H II region population, but N10 will have suffered higher extinction than
M17 owing to its greater distance (4.9 vs 1.6 kpc) and also because it is inside
the galactic molecular ring. 

Overall the comparison with M17 seems quite fair and M17 should not be too 
heavily extincted by intervening unrelated gas in the MSX bands.  However, 
given that all 
of the sources in the galactic center (not just those in Sgr B2) probably 
suffer from some significant foreground extinction, the similarity of their
spectra to that of M17 imples that the extinction must be rather gray, as
previously indicated by the individual IRDC discussed in Sect. 5 and
and illustrated in Figs. 8 and 9 (see \cite{IndMat+05} and \cite{NisNag+08}).

\begin{figure}
\includegraphics[height=14cm]{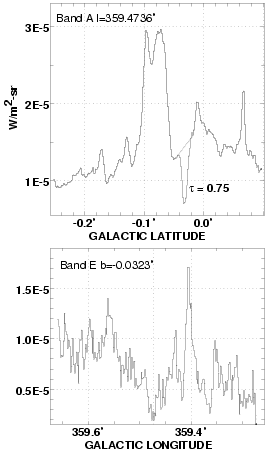}
\caption[]{MSX intensity (W m$^{-2}$ sr$^{-1}$) along spatial strips through 
the IRDC near Sgr C.  Top:  a strip in galactic latitude at l = 359.4736\degr
in MSX Band A at 8.2$\mu$. When the intensity distribution is interpolated
across the IRDC as shown the optical depth is 0.75 (0.815 mag).  Bottom:
a strip in longitude in MSX Band E around 22$\mu$.  Spatial resolution
is 21\arcsec\ = 0.00583\degr.} 
\end{figure}

\subsection{Sgr B at $l > 0.6$\degr, Sgr B2}

Spectra of compact sources in Sgr B  are shown in the righthand panels of 
Fig. 10.  The systematics of spectra from the more heavily-extincted 
sources in Sgr B2 at $l > 0.6$\degr\ in the lower right panel seem 
especially clear: the IR spectra steepen noticeably as the sources 
brighten and the $22\mu$/radio flux ratio increases.  For one of the
sources (\#7), and in particular that source which is closest to Sgr B1, the
IR/radio flux ratio actually reverses relative to M17, which is a
characteristic of nearly all of objects in Sgr B1.  

This steepening belies the similarity with M17 and N10 seen in the large in 
the upper left panel of Fig. 10 and complicates both the IR/radio flux 
comparison and the inference that the IR extinction is gray.  To flatten 
the M17 or N10 spectra across the MSX bands would require some 17 or 5 times 
higher extinction at 22$\mu$ than at $8\mu$, respectively.  To ameliorate 
this situation, we suggest that the observing geometry is at least partially 
responsible. The weakest Sgr B2 IR source shown is probably absent in the 
MSX data and the flat-spectrum measured flux is a mixture of noise and 
foreground haze (in the discussion
of Sgr E in Sect. 6.3 we note that IR sources which lack radio emission 
entirely have flat spectra across the MSX bands).  Put another way, the 
emitting and extincting material are mixed and detailed models will have
to be constructed to account for the variations in flux and slope.



The weakest MSX source has the flattest IR spectrum and the 
largest radio/IR disparity; its MSX E-band/VLA 18cm flux ratio is low 
by a factor about 50 compared to the template spectrum of M17, 
corresponding to an additional 4 mag of IR extinction.  As the MSX 
fluxes increase in Sgr B2 (\#7) the IR spectrum steepens,  but for the 
brightest IR source the 18cm flux is unexpectedly small; this source
is an interloper and properly belongs to Sgr B1.   This
pattern is mostly repeated in Sgr B1 in the upper right hand
panel of Fig. 10 where again the two brightest MSX sources have
comparatively small 18cm flux.  Apparently, the brighter MSX E-band 
sources in Sgr B tend to have steeper IR spectra and higher 
Band E/18cm flux ratios. 

\subsection{Sgr B at $l < 0.5$\degr, Sgr B1}

Spectra of sources in and just West of Sgr B1 at $l \la 0.5$\degr\ are 
shown at upper right in Fig. 10.   Note that the Sgr B1 sources are 
brighter in the IR than those Sgr B2, despite the fact that Sgr B2 is 
known as  the region of greater star formation activity judged from 
the radio 
emission.   The IR spectra of the galactic center sources in the upper 
right panel of Fig. 10 resembles those of M17 or N10, none being as 
flat as was seen in Sgr B2.  However, the radio/IR flux ratios differ 
notably from M17, N10 and Sgr B2, being rather smaller.  Some of 
this variation in radio/IR flux ratio can probably be explained by 
optical depth effects in the radiocontinuum, if the optical depth is 
appreciable 

\subsection{Sgr E at $l < 1.1$\degr}

Spectra of sources in Sgr E, which is largely devoid of extended flux 
in the radiocontinuum, are shown at lower left in Fig. 10. IR spectra of
he Sgr E sources generally resemble M17 and those in Sgr B1, and
the radio/IR flux ratios are also much like those in Sgr B1 and unlike 
those in Sgr B2.  The Sgr E sources are brighter in the IR and weaker in the 
radio than those in Sgr B2.  

IR spectra were also extracted for three sources (nos 7-9, shown dashed and in 
blue) which show no obvious 18cm radiocontinuum at all; the radio fluxes 
reported for these sources are the totals within the boundaries of the 
respective IR images.  These sources are characterized by flatter MSX 
spectra.

\begin{figure}
\includegraphics[height=12cm]{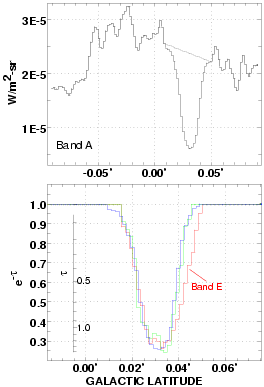}
\caption[]{MSX intensities (W m$^{-2}$ sr$^{-1}$) on a spatial strip 
through the IRDC near l=0.25\degr, running from (l,b) = 
(0.4\degr,-0.1\degr) to (0.2\degr,0.1\degr).  At top is the
strip in Band A, as in Fig. 11 and at bottom are the absorption
optical depth profiles in MSX bands A, C and E.  Strip coordinates
are labelled with the galactic latitude at positions along it
(the arc length is sqrt(2) larger than the latitude pixel spacing)
and the resolution is 21 sqrt(2)\arcsec\ = 0.00825\degr.}
\end{figure}


\section{Is there a CMZ in the Milky Way, or only a NSFR?}

How then is the gaseous material in the inner $\pm 175$ pc actually distributed?  
From our vantage point so near the mid-plane the 8$\mu$ light distribution in 
Fig. 1
is largely undifferentiated while the 22$\mu$ emission is at the other
extreme, showing the seemingly isolated Sgr B, A, and C source complexes. 
However, the \coth\ distribution in Fig. 1 is somewhat hollow and loop-like
at negative longitudes and the kinematics support this impression.  Even
from the crude sampling of recombination line emission in Fig. B.1 it is apparent
that the profiles are narrow compared to the span of the rotational pattern
across the central region.  Although kinematic projection effects are 
important, this is the signature of a hollow, ring-like distribution, not 
a filled disk.

\begin{figure*}
\includegraphics[height=13.2cm]{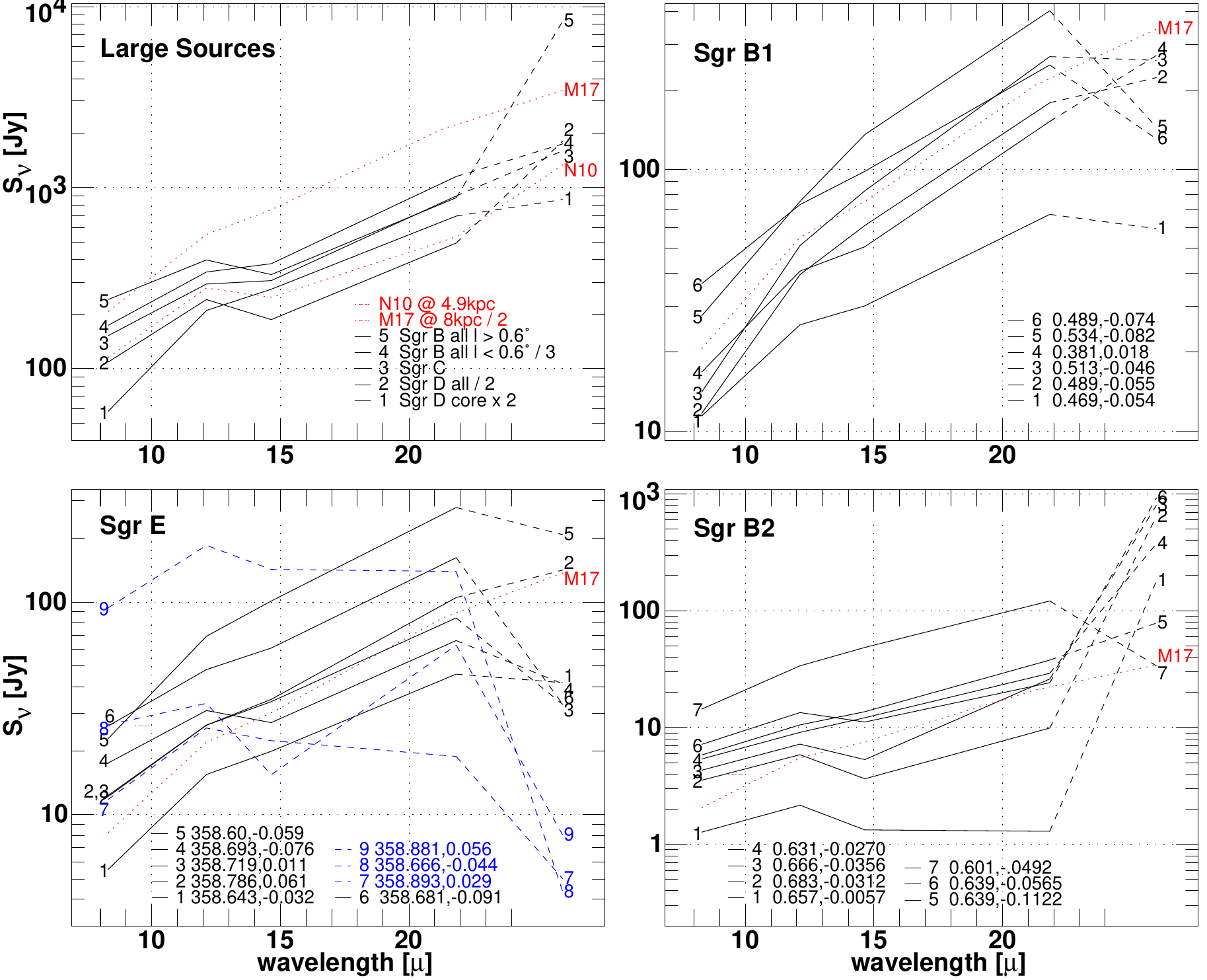}
\caption[]{MSX-VLA, IR-radio spectra of galactic center sources. 
18cm radio fluxes scaled upward by a factor 250 are shown at the
extreme right side of each curve. The global spectrum of M17 
\citep{PovSto+07} is shown in each panel as a template, 
arbitrarily scaled except at the upper left. Upper left: 
Integrated fluxes over extended regions.  In this panel the
M17 fluxes have been scaled to 8 kpc distance and divided by 2.   
Upper right: compact sources in Sgr B1.  Lower left: compact sources 
in Sgr E.  Lower right, compact sources in Sgr B2.}
\end{figure*}

Shown in Fig. 11 are two collapsed longitude-velocity maps of the 
\coth\ emission (see Fig 1), where in each panel the emission has 
been averaged either above or below the galactic equator (see also 
\cite{Sof95}). Emission associated with the Sgr source complexes generally 
appears in one of two kinematic ``branches'' which are manifested as 
nearly straight lines in the l-v plane, with similar velocity gradient 
(implying similar galactocentric radii) but very different longitude 
crossings at zero-velocity (implying a substantial component of 
non-circular motion, approximately 50 \kms).  The branch at positive 
latitude seen in the upper panel is displaced to 
negative velocity at l=0\degr, crossing 0-velocity at positive longitude 
and including the locus of the Sgr B2 recombination lines at l $\ga$ 
0.6\degr, v = 60 \kms.  Another branch at negative latitude seen in
the lower panel displays roughly symmetric behaviour, being displaced 
to positive velocity, crossing 0-velocity at negative longitude and 
meeting the locus of the Sgr C recombination lines at l=-0.5\degr, 
v = -60 \kms.  

An economical description of this behaviour puts the gas into a ring (or
 pseudo-ring comprised of arm segments which is 
slightly tipped out of the galactic plane, so that the front and back 
portions are displaced in latitude.  The outer size of the ring 
and its rotation speed are set by the locus of Sgr E at 
l=-1.25\degr\ corresponding to 175 pc, with a rotation speed of 220 
\kms\ (since the ring is viewed tangentially).  The line of sight
inclination is small, 3.2\degr, i.e. a vertical displacment of 
$\pm0.07$\degr\ across a diameter of 350 pc .  The near and far sides are 
also separated kinematically by a substantial ($\pm40-50$\kms) 
component of non-circular motion, perhaps indicating an elliptical 
shape to the ring. The ring in the Milky Way bar is somewhat on the small 
side, but not impossibly so,

\begin{figure*}
\includegraphics[height=9.9cm]{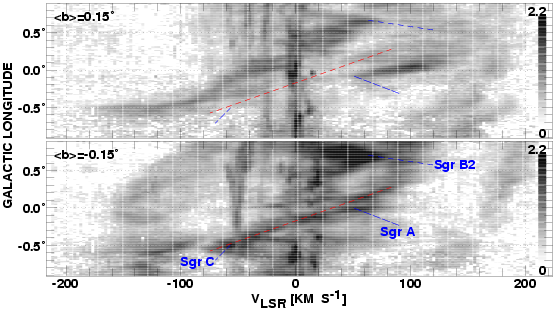}
  \caption{\coth\ Kinematics of the star-forming ring.  Shown are longitude-
velocity diagrams of \coth\ emission, where at each longitude the emission
has been averaged vertically, at 0\degr\ $<$ b $<$ 0.3\degr\ at top
and over -0.3\degr\ $<$ b $<$ 0\degr\ at bottom.  The locii of several 
H II regions and the locus of the lower-latitude kinematic branch 
of the ring are marked in the bottom panel.  Some of this annotation 
has been carried over to the top panel.}
\end{figure*}

If it can be decided which of the kinematic branches is the near side, 
the sense of the non-circular motion and the inclination follow directly 
and a good description of the geometry and kinematics can be made. 
In the context of the discussion here, the prominence and placement 
of the IR extinction at $l > 0.2$\degr\ suggest that the near branch 
of the ring is that which is seen at higher latitude and which includes 
Sgr B: Sgr C is then located on the ring portion seen behind the center.  
The same assignment of the nearer segment of the ring was noted earlier 
(Sect. 3.1) based on the shape of the global 8$\mu$ light distribution.
This is the same assignment made by \cite{Sof95} and repeated by
\cite{SawHas+04} based on the kinematics of molecular absorption lines.
It accounts for the presence of absorption at -100 \kms\ in the
vicinity of Sgr C, even if gas within the Sgr C complex at like velocities
is not responsible (see Sect. 3.3).

Nonetheless the ring description is incomplete inasmuch as it begs the question 
of the gas in Sgr A, which is believed to reside near the center, but 
coincidentally occurs in a velocity range (20-70 \kms) which is 
manifestly present along the far side of the ring around l=0\degr\ 
(see the lower panel of Fig. 1; \cite{Sof95} assigned the 50 \kms\
gas around Sgr A to a separate arm).  Given the appearance of the rings 
in other galaxies (which do not support the idea that the ring in the
Milky Way would fortuitously have a huge gap just behind Sgr A) it seems 
unavoidable that some portion of the material which is presently ascribed 
to Sgr A and the vicinity of the center must instead arise in the ring.  

Deciding which gas in Sgr A is actually central, and how much dense gas 
exists interior to the ring at all, are challenges for the future.  But if 
the Milky Way is like other galaxies the region interior to the ring 
may be largely devoid of dense gas and the description of the material 
as residing in a "central molecular zone" may be somewhat misleading.  
The molecular gas distribution presents a solid face in \cotw\
(not so much even in \coth, see Fig. 1) from our vantage point in 
the galactic plane, but it is only the thin annular region of the ring which
is actually occupied by the dense, warm, turbulent molecular gas which
observationally comprises the CMZ.

\section{Summary}

Nuclear star-forming rings in barred galaxies are ongoing starbursts,
fed by inflow of material along the dust lanes \citep{KorCor04}.  The 
inward flow is typically staunched by resonant phenomena 
\citep{RegTeu03,RegTeu04} at a narrow ring (of radius 100 - 1000 pc)
where the continued introduction of new gas at high velocity keeps the 
density and pressure high, and where the continuing but episodic
\citep{Lis06Shred} interaction of fresh and 
previously-accumulated gas induces rapid star formation through
such mechanisms as cloud-cloud collisions \citep{HasAra+08}.

The galactic center H II regions and molecular gas and their kinematics 
are not always discussed together but observational descriptions of a 
kinematic ring can be found in \cite{LisBur+77}, \cite{Sof95}, 
\cite{OkaHas+96} and \cite{SawHas+04}.  Mechanisms of star formation 
in a local ring were also discussed by \cite{StaMar+04}, although without 
reference to the actual Sgr H II regions, and the connection between 
barring in the Milky Way and an inner ring are noted explicitly in 
our earlier work and by \cite{RodCom+06} and \cite{RodCom08}.

The presence of an approximately 200-pc radius ring naturally accounts 
for many of the properties which have been attributed to molecular gas 
in the inner region or central molecular zone \citep{MorSer96} of the 
Milky Way -- high mean density, thermal pressure and large turbulent
motion seen over a very extended region around the center -- but without
actually filling the interior volume, accomodating yet other, more 
diffuse constituents of the galactic center ISM such as the diffuse
metastable \H3\p\ observed by \cite{OkaGeb+05}. Dense molecular 
gas, especially in the Sgr A complex, may exist interior to the ring,
but conversely, some of the gas at 10-50 \kms\ seen against Sgr A is 
probably in the ring (see just below and Sect. 6).

In this work, we compared the observed distributions of the IR and (radio)
L-band continuum and molecular gas in order to describe the structure
and some of the processes at work in the Milky Way's nuclear ring. For
the processes, we noted (Sect. 3.2) that Sgr B2 has already been described 
as a cloud-cloud collision and we described the particular morphology
which led to this hypothesis, a high velocity cap projected against a broader
molecular gas shell at lower velocity surrounding the Sgr B 
H II regions.  We showed (Sect. 3.3) that Sgc C has similar kinematics, which
accounts for the puzzling brightening of CO at widely-separated velocities
in its direction. Sgr B and C are seen at very similar separations 
0.5\degr-0.6\degr\  from the center and have similar velocities.  In terms of 
bar dynamics, cloud-cloud collisions in Sgr B and C would represent the
contact points near the inner ends of the bar dust lanes where 
material already in the ring interacts with inflowing or sprayed gas.

The H II region complexes are all composite, but the apparent groupings
in Sgr D (Sect. 3.1) and Sgr E (3.4) may be unphysical.   The Sgr D 
H II region at $l = 1.1$\degr, v = -10 \kms\ cannot be proven to be
in the galactic center nor is there direct evidence that it and the
immediately adjacent (osculating, even), similarly-sized SNR are 
associated.  Unfortunately, searches for tracers which would reveal the 
systemic velocity of the remnant have been unsuccessful.  Sgr E seen
at the presumed negative longitude extremity of the ring at 
$l \la -1.2$ \degr, v $< -200$ \kms\ is certainly in the nuclear region
but the implied tangential viewing geometry will cause unrelated
and spatially-separated material to appear at similar velocity.  It
seems telling that Sgr E is a collection of discrete objects with
little extended flux or apparent structure and rather modest  H II
regions powered by B-stars, i.e. field objects.

The appearance of the galactic center in the 8-22$\mu$ IR MSX bands
is strongly affected by extinction and the MSX bands fall in something
of a plateau in the otherwise steady decline of extinction with increasing 
wavelength (Sect A.1, Fig. A.1).  We began by noting (Sect. 3, Fig. 1) 
that the large-scale distribution of $8\mu$ light is bowed because it is 
preferentially extincted at its northern edge and toward its extremities.
This immediately suggested a ring distribution of the extincting material
with a slight upward inclination of the near side.  The presence of such
a tilt in the ring is also seen kinematically because the front and back
ring portions have a systematic separation in latitude and velocity at
most longitudes (see below and Sect 6).   

We extracted the apparent extinction of prominent IR-absorbing
material near Sgr C and between Sgr A and B.  Especially in the latter 
the depth of the extinction ($\tau = 1.35$) does not vary noticeably over 
the MSX bands, indicating that the extinction generally is indeed gray.  
Another indication that
the extinction is gray comes from the overall similarity of IR spectra
of galactic center and galactic disk H II regions, because the former are
much more heavily extincted but are not seen to be reddenend.

We also extracted the $8-22\mu$ spectra of the H II region complexes and
their constituents (Sect. 5 and Fig. 10), and showed that globally, the 
spectra and apparent luminosities of galactic center sources are like 
those of disk objects such as M17 and N10.  However, there is also evidence 
of 
strong IR extinction, because the ratio of radio L-band (18cm) to IR 
flux is much larger in Sgr B and somewhat larger even in Sgr D.  Within
Sgr B, the extinction of Sgr B2 at $l > 0.6$\degr\ behind the 
high-velocity cap signifying the cloud collision is so great that
the strongest radio H II regions are simply absent in the MSX maps.
Overall, the galactic center sources resemble M17 or N10 in their
IR spectra but there is considerable variation of the radio/IR flux
ratio even when extinction cannot obviously be fingered as the 
culprit.  This probably calls for detailed modelling.

Last, we discussed the ring morphology and kinematics (Sect. 6).  
We discussed how
the kinematic patterns of the CO emission fall into narrow-lined
branches within the overall velocity envelope (i.e. are ring like)
 such that the gas, although mostly rotating at 200-220 \kms, 
also has 30-50 \kms\ non-circular motions which cause the near and
far portions to appear at different velocity at the same longitude.
One portion of the ring therefore shows substantial positive velocities 
over the longitude range of the Sgr A complex, where much of the molecular
emission usually attributed to the immediate vicinity of the center
has always, to some general puzzlement, appeared at v = 20-40 \kms.
Some of this material probably orginates in the ring, well away from the 
center, because there is no obvious reason to expect a complete gap in
the ring over the approximately 0.2\degr\ extent of the Sgr A complex.

The ring portions are separated in velocity and displaced in 
latitude, consistent with the first impression gained from the bowing
of the overall distribution of IR light in the presence of prominent
extinction, whereby the front side of the ring is inclined slightly
upward (Sect. 3).  Viewed in detail (Fig. 11) , the blue-shifted branch
of the ring gas appears at higher latitude
and intersects the velocity of Sgr B2 H II recombination
line emission at positive longitude, 60 \kms\ at $l \ga 0.6$\degr.
The red-shifted branch at lower latitude serves a similar role for
Sgr C (v=-60\kms at l = -0.5\degr) and includes the positive-velocities
present across  the region of the Sgr A complex.  Therefore, Sgr B and Sgr
C are on opposite front-back portions of the ring and Sgr B, which
lies in the higher-latitude branch ( responsible for
the prominent extinction), falls on the near side, as has generally
been assigned in the past based on molecular absorption line measurements.

Maps of molecular emission and absorption have now shown at least a 
partial view of many of the stages of the inflow and circulation of 
bar material, from individual uptake events well outside the center 
through the bar dust lanes into and within the nuclear star-forming 
ring.  Our unparalleled viewpoint within the galactic disk and near
the ring midplane provides a unique, if often bewildering perspective;
all that remains is to take advantage of it.

\begin{acknowledgements}

The NRAO is operated by AUI, Inc. under a cooperative agreement 
with the US National Science Foundation. The Kitt Peak 12-m millimetre 
wave telescope is now operated by the Arizona Radio Observatory (ARO), 
Steward Observatory, University of Arizona.  

\end{acknowledgements}

\appendix{}

\section{MSX bands}

The MSX bands and wavelengths are enumerated in Table 1 of the text.  
Figure A.1 shows how these bands overlap the extinction curves for galactic dust
with \RV\ = 3.1 and 5.5 \citep{WeiDra01}.

\begin{figure}
\includegraphics[height=7.7cm]{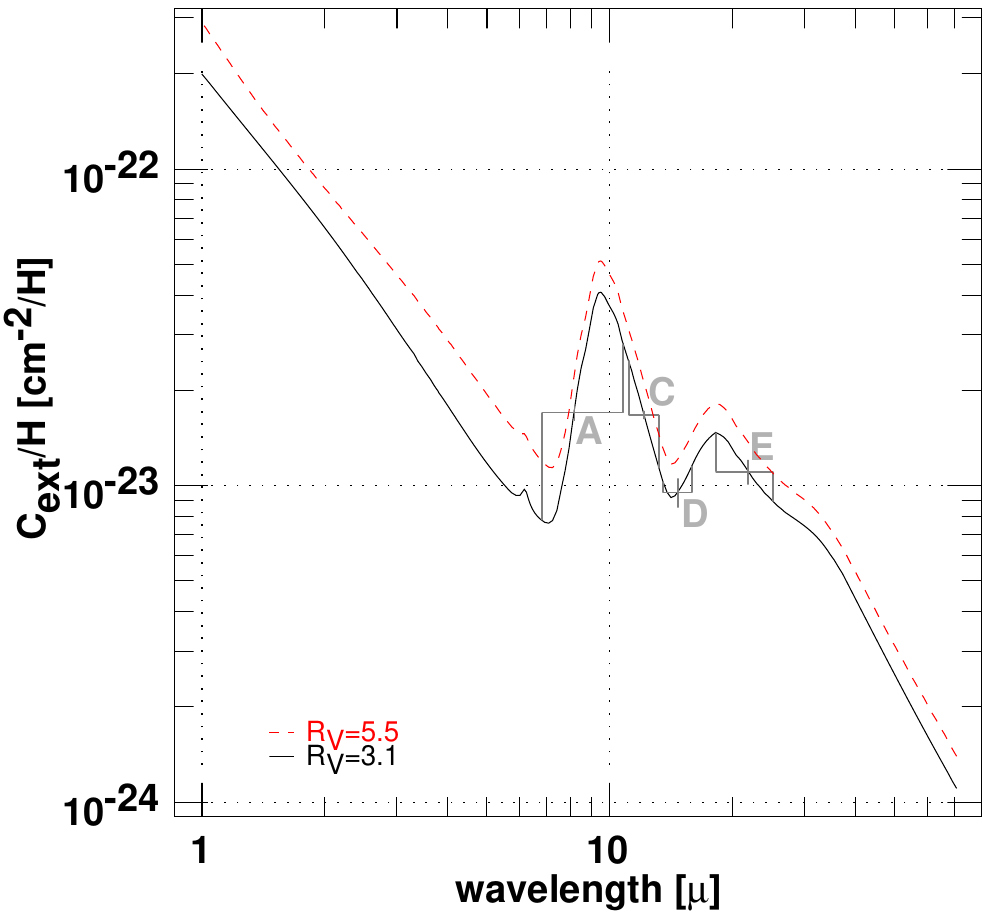}
\caption[]{Extinction cross-section per H-particle for galactic 
dust with \RV\ = 3.1 and \RV\ = 5.5 from \cite{WeiDra01}.}
\end{figure}

\section{Recombination Line Velocities}

19-GHz H70$\alpha$ recombination line profiles drawn from the work of
\cite{Lis92} are shown in Fig. B.1.

\begin{figure*}
\includegraphics[height=15.5cm]{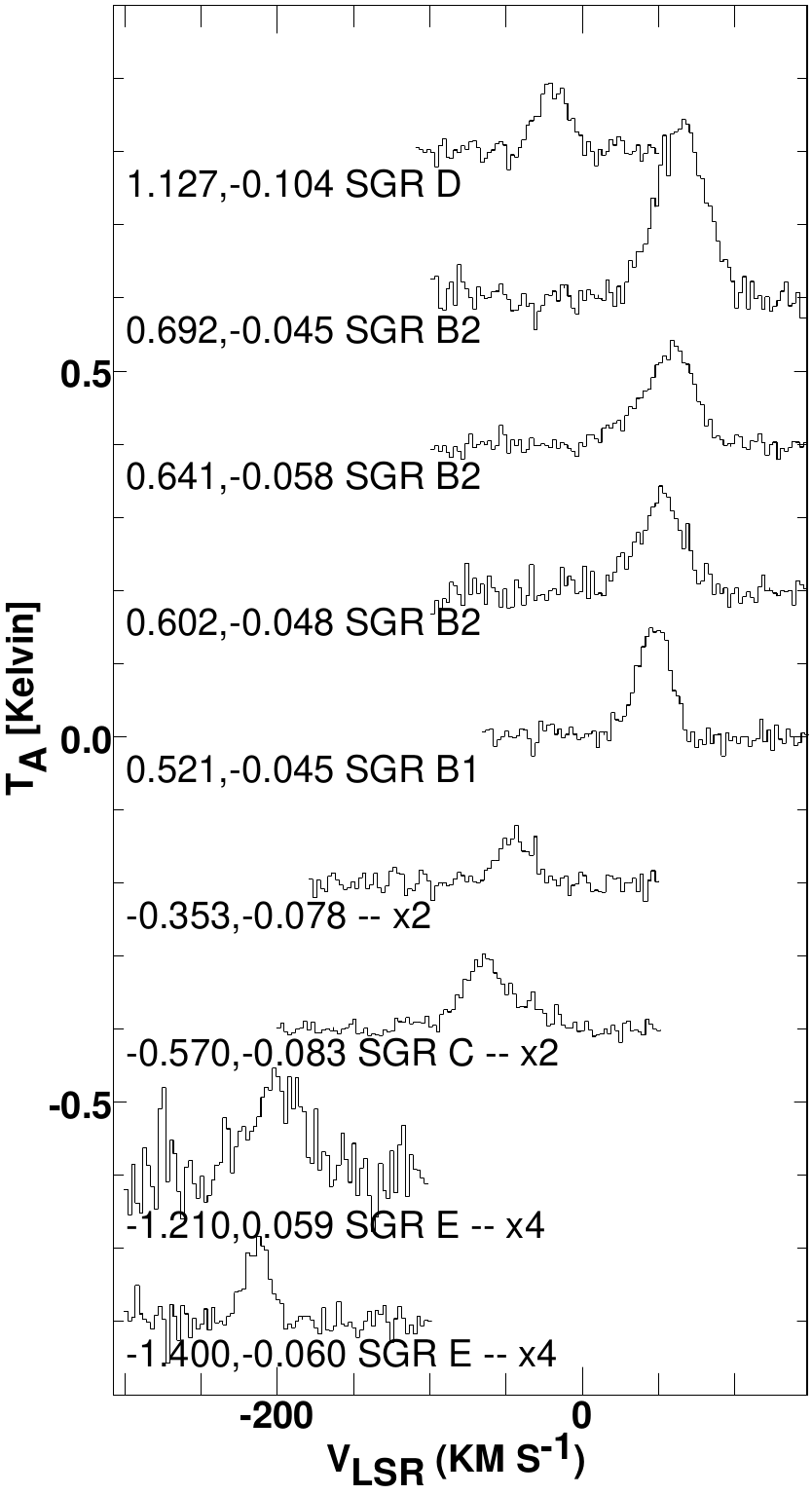}
\caption[]{19-GHz H70$\alpha$ radio recombination line velocities toward
H II regions seen over a range of longitudes within the star-forming ring 
\citep{Lis92}.}
\end{figure*}

\section{N10: An exemplary H II region bubble}

N10, at a kinematic distance of 4.9 kpc is one of three exemplary H II region 
bubbles recently discussed by \cite{WatPov+08}.  Its resemblance to Sgr C
is obvious.

\begin{figure*}
\includegraphics[height=13cm]{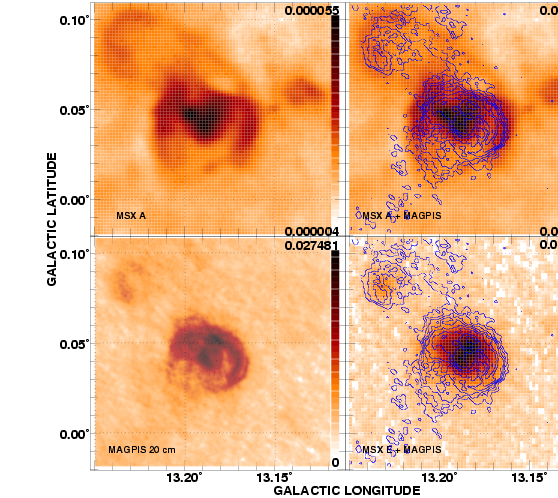}
\caption[]{MSX and MAGPIS 20cm images of N10. Upper left: MSX A band.  Lower left:
MAGPIS 20cm image \citep{HelBec+06}. Upper right:  
contours of 20cm radiocontinuum emission
overlaid on the MSX A band image.  Lower left: same for MSX E band.
See also \cite{WatPov+08}. }
\end{figure*}

\bibliographystyle{apj}

\end{document}